\documentclass[submission,copyright]{eptcs}
\usepackage{breakurl}             % Not needed if you use pdflatex only.
\usepackage[utf8x]{inputenc}
\usepackage{booktabs}
\usepackage{graphicx}
\usepackage{paralist}
\usepackage{xspace}
\usepackage{hyperref}
\usepackage{comment}
\usepackage{paralist}
\usepackage{cite}
\usepackage{xurl}
\usepackage{xstring}
\usepackage{textcase}
\usepackage{mfirstuc}
\usepackage{ragged2e}
\usepackage{svg}
\usepackage[htt]{hyphenat}
\usepackage{amsmath,amssymb}
\usepackage[flushleft]{threeparttable}
\usepackage[inline]{enumitem}
\usepackage{multirow}
\usepackage{subcaption}
\usepackage[normalem]{ulem}
\usepackage{todonotes}
\setlist[enumerate,1]{label=\textit{\alph*)}}
\hypersetup{
	colorlinks = true, % false: boxed links; true: colored links
	linkcolor=black, % color of internal links
	citecolor=black, % color of links to bibliography
	urlcolor=black % color of external links
}
\usepackage{tikz}
\usetikzlibrary{automata,positioning}
\usepackage{times}
\usepackage[]{listings}
\lstdefinelanguage{AsmetaL} %%AsmetaL
{morekeywords={module, par, endpar, if, endif, then, else, seq, endseq,
		signature, definitions, asm, import, export, function, domain, main, rule, macro,
		invariant, over, choose, let, endlet, with, ifnone, forall, abstract, default,
		init, do, agent, dynamic, controlled, monitored, in, out, static, derived,
		subsetof, switch, case, endswitch, enum, CTLSPEC, LTLSPEC, JUSTICE, mod},
	sensitive=true, morecomment=[l]{//}, morecomment=[s]{/*}{*/},
	morecomment=[l][\color{white}\tiny]{'},%for creating empty rows at the end
	morestring=[b]",tabsize=2, columns=fullflexible, basicstyle=\scriptsize, captionpos=b,escapechar=|}

\lstdefinelanguage{Avalla} %%Avalla
{morekeywords={scenario, load, invariant, set, exec, step, until, check, execblock, begin, end, library},
	sensitive=true, morecomment=[l]{//}, morecomment=[s]{/*}{*/},
	morecomment=[l][\color{white}\tiny]{'},%for creating empty rows at the end
	morestring=[b]",tabsize=2,columns=fullflexible,showlines=true, basicstyle=\scriptsize}

\colorlet{punct}{red!60!black}
\definecolor{background}{HTML}{EEEEEE}
\definecolor{delim}{RGB}{20,105,176}
\colorlet{numb}{magenta!60!black}

\lstdefinelanguage{json}{
	basicstyle=\scriptsize,
	showstringspaces=false,
	breaklines=true,
	frame=lines,
	literate=
	*{0}{{{\color{numb}0}}}{1}
	{1}{{{\color{numb}1}}}{1}
	{2}{{{\color{numb}2}}}{1}
	{3}{{{\color{numb}3}}}{1}
	{4}{{{\color{numb}4}}}{1}
	{5}{{{\color{numb}5}}}{1}
	{6}{{{\color{numb}6}}}{1}
	{7}{{{\color{numb}7}}}{1}
	{8}{{{\color{numb}8}}}{1}
	{9}{{{\color{numb}9}}}{1}
	{:}{{{\color{punct}{:}}}}{1}
	{,}{{{\color{punct}{,}}}}{1}
	{\{}{{{\color{delim}{\{}}}}{1}
	{\}}{{{\color{delim}{\}}}}}{1}
	{[}{{{\color{delim}{[}}}}{1}
	{]}{{{\color{delim}{]}}}}{1},
}

\newcommand{\asmtocode}{\texttt{Asm2C{\small++}}\xspace}
\newcommand{\asmeta}{ASMETA\xspace}
\newcommand{\asmetaL}{\texttt{AsmetaL}\xspace}

\newcommand{\avalla}{{\tt Avalla}\xspace}
\newcommand{\cpp}{C\nolinebreak\hspace{-.05em}\raisebox{.2ex}{\small\bf ++}\xspace}

\newcommand{\imagePath}{images/}

\newcommand{\code}[1]{{\small \sf {#1}}}

\title{Developing a Prototype of a Mechanical Ventilator Controller from Requirements to Code with \asmeta}

\author{Andrea Bombarda
	\institute{University of Bergamo, Bergamo, Italy}
	\email{andrea.bombarda@unibg.it}
	\and
	Silvia Bonfanti
	\institute{University of Bergamo, Bergamo, Italy}
	\email{silvia.bonfanti@unibg.it}
	\and
	Angelo Gargantini
	\institute{University of Bergamo, Bergamo, Italy}
	\email{angelo.gargantini@unibg.it}
	\and
	Elvinia Riccobene
	\institute{University of Milano, Milan, Italy}
	\email{elvinia.riccobene@unimi.it}
}

\providecommand{\event}{AppFM 2021}
\def\titlerunning{Developing a Prototype of a Mechanical Ventilator Controller from Reqs to Code with \asmeta}
\def\authorrunning{A. Bombarda, S. Bonfanti, A. Gargantini \& E. Riccobene}

\begin{document}
	\maketitle

\begin{abstract}
Rigorous development processes aim to be effective in developing critical systems, especially if failures can have catastrophic consequences for humans and the environment.
Such processes generally rely on formal methods, which can guarantee, thanks to their mathematical foundation, model preciseness, and properties assurance. However, they are rarely adopted in practice.

In this paper, we report our experience in using the Abstract State Machine formal method and the \asmeta framework in developing a prototype of the control software of the MVM (Mechanical Ventilator Milano), a mechanical lung ventilator that has been designed, successfully certified, and deployed during the COVID-19 pandemic.
Due to time constraints and lack of skills, no formal method was applied for the MVM project. However, we here want to assess the feasibility of developing (part of) the ventilator by using a formal method-based approach.

Our development process starts from a high-level formal specification of the system to describe the MVM main operation modes. Then, through a sequence of refined models, all the other requirements are captured, up to a level in which a C++ implementation of a prototype of the MVM controller is automatically generated from the model, and tested. Along the process, at each refinement level, different model validation and verification activities are performed, and each refined model is proved to be a correct refinement of the previous level.
By means of the MVM case study, we evaluate the effectiveness and usability of our formal approach.
\end{abstract}

\section{Introduction}
To prevent catastrophic consequences for humans and the environment due to system failure or unsafe operation, safety-critical software requires development methods and processes that could lead to provably correct system operation~\cite{Lutz00,Leveson20}. From long time the use of models and formal analysis techniques is highly demanded already at design-time to improve software quality and guarantee safety, reliability, and other desired qualities. However, the usage of formal methods in industrial projects is still limited~\cite{10.1145/1134285.1134406,Gleirscher2020,Garavel2020} and practitioners are still 	skeptical in using formal methods since they are considered time-consuming approaches that do not fit into an agile continuous integration development process.

Besides the well-known lack of training, among the barriers to the adoption of formal methods, we can remark (i) the complexity of formal notations, (ii) the poor scalability, (iii) the lack of easy-to-use tools supporting modeling, validation, and verification activities at the design phase, and (iv) the gap between models and code. Formal approaches allowing model refinement would help the designer in facing the complexity of system requirements, and techniques of automatic code generation from models would allow for producing correct-by-construction code/artifacts  of  the  system in a seamless manner from the requirements to the final implementation.
We believe that some characteristics of formal methods would also favor their use in the assurance process:
models should possibly be executable for high-level design validation and endowed with properties verification mechanisms; operational approaches are more adequate than denotational ones to support (automatic) code generation from models and model-based testing.

In principle, different methods and tools can be used to guarantee software safety and reliability; however, the integrated use of different tools around the same formal method is much more convenient than having different tools working on input models with their own languages.

Among the plethora of existing formal methods, the Abstract State Machines (ASMs)~\cite{Boerger2003,Boerger2018} are a system engineering method that can guide the development of software systems seamlessly from requirements capture to their implementation. This is shown by the adoption of ASMs in a series of cases studies, as in \cite{Bombarda2019,LGSjournalSTTT2017,abz2016siSCICO2017,ArcainiABZ2020},  to name a few. Although ASMs have a rigorous mathematical foundation -- as transition systems that extend the Finite State Machines (FSMs)~\cite{Boerger2003}--, ASMs can be understood as pseudo-code or virtual machines working over abstract data structures. Besides their \emph{pseudo-code format},
\begin{inparaenum}[(1)]
\item   ASM models can be specified at any desired \emph{level of abstraction} and are \emph{executable models};
\item \emph{model refinement} is an embedded concept in the ASM formal approach; it allows facing the complexity of system specification by starting with a high-level description of the system and then proceeding step-by-step by adding further details till a desired level of specification has been reached; each refined model must be proved to be a correct refinement of the previous one, and checking of such relation can be performed automatically~\cite{Arcaini2016Vis};
\item the concept of ASM \emph{module}, i.e., an ASM without the main firing rule, facilitates model scalability and separation of concerns, so tackling the complexity of big systems specification;

\item ASM-based modeling and analysis are supported by a set of tools that can be used in an integrated manner within the \asmeta (ASM mETAmodeling)~\cite{modelDrivenProcess,asmetaWeb,Arcaini2021} framework.
\asmeta provides tools for specifying the executable behavior of a system, for checking properties of interest, specifying and executing validation scenarios, generating prototype code, etc.
\end{inparaenum}

During the COVID-19 pandemic, our research team was involved in the design, development, and certification of a mechanical lung ventilator called MVM (Mechanical Ventilator Milano)\footnote{\url{https://mvm.care/}} \cite{Abba2021}. The project started from an idea of the physicist Cristiano Galbiati, who was soon joined by dozens of physicists, engineers, physicians, and computer scientists from 12 countries around the world, including the authors of this paper.
The team was able to realize a ventilator that is reliable, easily reproducible on a large scale, available in a short amount of time, and at a limited cost~\cite{Guardo2021}. The MVM has obtained the FDA (Food and Drug Administration) Emergency Use Authorization (EUA) followed by similar authorizations issued by Health Canada and the CE marking as well.
During the development of the software, no formal method has been applied mainly because of time constraints and a lack of developers' skills with any formal method. However, we wanted to assess the feasibility (and possibly also the limits) of developing (part of) the ventilator by using a formal method-based approach.
Because we were involved in all the software development process and its certification (we wrote the requirements and were involved also in the coding and testing for the MVM), we have all the knowledge and expertise necessary to perform this experiment. In this paper, we report the practical experience of using ASMs/\asmeta in modeling, analyzing, and encoding the control software of the MVM.

The paper is structured as follows. Sect.~\ref{sec:asmeta} presents the background regarding the \asmeta framework and Sect.~\ref{sec:MVM} introduces the MVM case study. In Sect.~\ref{sec:vev} we present the modeling, verification, and validation activities performed for the case study.
The automatic generation of \cpp code and unit tests from the ASMETA specification is described in Sect.~\ref{sec:codegen}, while the deployment process on Arduino is described in Sect.~\ref{sec:arduino}.
Sect.~\ref{sec:discussion} discusses the pros and cons of the ASMETA approach and the aspects that could prevent or favor its use in practice and concludes the paper.

\section{Abstract State Machines and \asmeta development process}\label{sec:asmeta}
The formal design of the MVM we propose here is based on the ASMs formal method~\cite{Boerger2003,Boerger2018}, which is an extension of FSMs where unstructured control states are replaced by states with arbitrarily complex data. The development process from formal requirement specification to code generation has been supported by the \asmeta framework~\cite{Arcaini2021}, a set of tools around the ASMs, which we used for modeling the ventilator and performing validation and verification activities, together with automatic source code and tests generation.

ASM \emph{states} are mathematical structures, i.e., domains of objects with functions and predicates (i.e., boolean functions) defined on them. The \emph{transition} from one state $s_i$ to the next state $s_{i + 1}$ is obtained by firing the set of all ASM \emph{transition rules} 
invoked by a unique \emph{main} rule, which is the starting point of a computation step. Transition rules express the modification of dynamic controlled functions interpretation from one state to the next one. Indeed, functions are classified as \emph{static} (never change during any run of the machine) or \emph{dynamic} (may change as a consequence of agent actions or \emph{updates}). Dynamic functions are distinguished between \emph{monitored} (only read by the machine and modified by the environment) and \emph{controlled} (read in the current state and updated by the machine in the next state).  \emph{Derived} functions are not part of the state since they are defined in terms of other (dynamic) functions. 

The \emph{update} rule, as assignment of the form $\mathit{f(t_1,\ldots, t_n)} := v$, is the basic unit of rules construction, being $f$ an n-ary function,  $t_i$ terms, and $v$ the new value of $\mathit{f(t_1,\ldots, t_n)}$ in the next state.
By a limited but powerful set of \emph{rule constructors}, function updates can be combined to express other forms of machine actions as, for example, guarded actions (\texttt{if-then}) and simultaneous parallel actions (\texttt{par}).

\begin{figure}[t]
	\centering
	\includegraphics[width=0.75\linewidth]{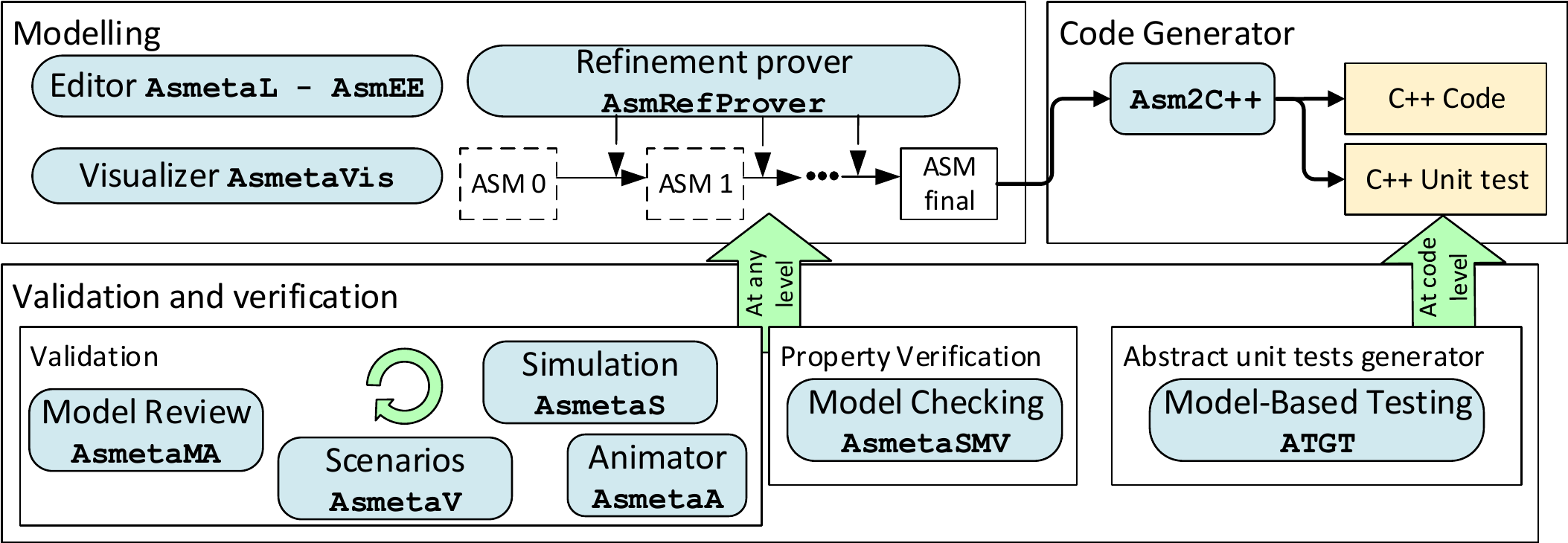}
	\caption{The ASM development process powered by the \asmeta framework}
	\label{fig:ASMProcess}
\end{figure}

The ASM method can facilitate the entire life cycle of software development,
i.e.,  from modeling to code generation. Fig.~\ref{fig:ASMProcess} shows the
development process based on ASMs and supported by the \asmeta framework\footnote{\url{https://asmeta.github.io/}}~\cite{modelDrivenProcess}
which provides a set of tools to help the developer in the following
activities: modeling, validation, verification, and, when required, code
generation.

In the modeling phase, the user specifies the system models by using the
{\tt AsmetaL} language, and
the editor {\tt AsmetaXt} provides some useful
editing support. ASM models can be read as pseudo-code over abstract data types.
Moreover, the ASM visualizer {\tt AsmetaVis} can be used to transform the textual model into graphs using the ASM graphical notation proposed in~\cite{Arcaini2016Vis}. The refinement correctness can be automatically proved using the tool {\tt ASMRefProver}~\cite{arcaini2016Ref}.

The validation process is supported by the model simulator {\tt AsmetaS}, the animator {\tt AsmetaA}, the scenarios executor {\tt AsmetaV} -- all supporting models exposing time features~\cite{Bombarda2021} --, and the model reviewer {\tt AsmetaMA}. The simulator {\tt AsmetaS} allows performing two types of simulation: interactive simulation (the user inserts the value of monitored functions) and random simulation (the tool randomly chooses the value of monitored functions among those available). A 
tabular presentation of a simulation is possible by means of the animator {\tt AsmetaA} that shows the models' execution through the use of tables. {\tt AsmetaV} executes scenarios written in \avalla language. Each scenario contains the expected system behavior and the tool checks whether the machine runs correctly. 
The model reviewer {\tt AsmetaMA} performs the static analysis. It determines whether a model has sufficient quality attributes, e.g., minimality (the specification does not contain elements defined or declared in the model but never used), completeness (it requires that every behavior of the system is explicitly modeled), and consistency (it guarantees that locations are never simultaneously updated to different values).

Property verification is performed with the {\tt AsmetaSMV} tool. It verifies if the properties derived from the requirements are satisfied by the models. When a property is verified, it guarantees that the model complies with the intended behavior.  {\tt AsmetaSMV} exploits the NuSMV and NuXmv model checkers. Linear Temporal Logic (LTL) and Computation Tree Logic (CTL) properties can be proved by using the classical NuSMV model checker in the case of limited domains. NuXmv (an extension of NuSMV) is able to deal also with unlimited domains in the case of LTL properties and allows for proving properties based on \emph{time} values. 

In case the code is available, the \asmeta framework provides the {\tt ATGT} tool that generates abstract unit tests starting from the ASM specification by exploiting the counterexample generation of a model checker (NuXMV or NuSMV). If not, a tool that automatically generates \cpp code from ASMs is available ({\tt \asmtocode})~\cite{Bonfanti2019}.

\section{MVM Case study}\label{sec:MVM}

MVM~\cite{Abba2021} is an electro-mechanical ventilator equivalent to the old and reliable Manley Ventilator~\cite{manleyVentilator}. It is intended to provide ventilation support for patients that are in intensive therapy and that require mechanical ventilation.
MVM works in pressure-mode, i.e., the respiratory time cycle of the patient is controlled by the pressure, and, therefore, this ventilator requires a source of compressed oxygen and medical air that are readily available in intensive care units.  More precisely, MVM has two operative modes: \emph{Pressure Controlled Ventilation} (PCV) and \emph{Pressure Support Ventilation} (PSV).
In the PCV mode, the respiratory cycle is kept constant and the pressure level changes between the target inspiratory pressure and the positive end-expiratory pressure. New inspiration is initiated either after a breathing cycle is over, or when the patient spontaneously initiates a breath. In the former case, the breathing cycle is controlled by two parameters: the respiratory rate (RR) and the ratio between the inspiratory and expiratory times (I/R). In the latter case, a spontaneous breath is triggered when the MVM detects a sudden pressure drop within the trigger window during expiration.
The PSV mode is not suitable for patients that are not able to start breathing on their own because the respiratory cycle is controlled by the patient, and MVM partially takes over the work of breathing. A new respiratory cycle is initiated with the inspiratory phase, detected by the ventilator when a sudden pressure drop occurs. When the patient's inspiratory flow drops below a set fraction of the peak flow, MVM stops the pressure support, thus allowing exhalation. If a new inspiratory phase is not detected within a certain amount of time (apnea lag), MVM will automatically switch to the PCV mode because it is assumed that the patient is not able to breathe alone.

The ventilator allows the air to enter/exit through two valves, i.e., an input valve and an output valve. When the ventilator is not running, the valves are set to safe mode: input valve closed and output valve opened. When the inspiration starts, the input valve is opened and the output valve is closed, while during the expiration the input valve is closed and the output valve is opened.
Both in PCV and PSV mode inspiratory pause, expiratory pause, and recruitment manoeuvrer are allowed by user request. Inspiratory/Expiratory pause consists in closing the input and output valves of the ventilator respectively after the inspiration and expiration phase. The inspiratory pause allows measuring the pressure reached inside the alveoli at the end of the inspiratory cycle, while the expiratory pause allows measuring the residual pressure to check possible
%level of
obstruction in the exhalation channel. Recruitment manoeuvrer is an emergency procedure required after intubation and it consists in prolonged lung inflation as necessary to reactivate the alveoli immediately; during this manoeuvrer, the input valve is opened and the output valve is closed.

Before starting the ventilation the MVM controller passed through three phases. The \emph{start-up} in which the controller is initialized with default parameters, \emph{self-test} which ensures that the hardware is fully functional, and \emph{ventilation off} in which the controller is ready for ventilation when requested.
 
To give an idea of the complexity of the entire MVM, its detailed behavior is described in the requirements documents which count altogether about 1000 requirements, each being a brief sentence. One document describes the behavior of the overall system, while 15 requirements documents describe the detailed behavior of software components. The controller itself has its own requirement document which consists of 31 pages and 157 requirements.

\section{Modeling and V\&V}\label{sec:vev}
In this section, we present the modeling of the MVM controller and the validation and verification activities we have performed.
We proceeded through three refinement steps. \begin{inparaenum}[(1)]
\item   The first model (\code{MVMController00}) describes the transition between the main operation phases: startup, self-test, ventilation off, PCV, and PSV modes.
\item The second model (\code{MVMController01}) introduces the modeling of inspiration and expiration in both PCV and PSV,
\item while the third model (\code{MVMController02}) adds the expiratory/inspiratory pauses, the recruitment manoeuvrer, and the apnea.
\item The last refinement step (\code{MVMController03}) introduces (in both PCV and PSV) the transition between expiration and inspiration in case of pressure drop, and the transition between inspiration and expiration in case the pressure exceeds a threshold.
\end{inparaenum}
The time features have been modeled using the \code{TimeLibrary}~\cite{Bombarda2021}.

We have proved the refinement correctness by using the SMT-based tool {\tt Asm\-Ref\-Prover} which proves the stuttering refinement as defined in \cite{arcaini2016Ref}.

In order to have an idea of the complexity of the ventilator models, Table~\ref{tab:dimension} shows the models' dimensions in terms of the number of functions and rules.

\begin{figure*}[t]
\begin{minipage}[t]{0.54\textwidth}\vspace{0pt}%
	\scriptsize
	\begin{tabular}{lccccccc}
		\toprule
 declarations			\multirow{2}{*}{} & \multicolumn{4}{c}{\# Functions}                                                                                                & &\multicolumn{2}{c}{\# Rules}  \\
			\cmidrule{2-5}  \cmidrule{7-8}
		& mon. & con. & der. & static & & decl. & rules \\ \midrule
		TimeLibrary & 1  	& 2 & 2 & 0 & & 2 & 2 \\
		MVMController00      & 5                              & 1                              & 0                            & 0                           && 8     & 27                          \\
		MVMController01      & 6  & 5                              & 0                           & 5                           && 19                                        & 98                          \\
		MVMController02      & 9                             & 6                              & 0                           & 9                           && 27                                        & 160                          \\
		MVMController03      & 11                             & 6                              & 0                           & 10                           && 27                                        & 170                          \\ \bottomrule
	\end{tabular}
	\captionof{table}{Models dimension (including the TimeLibrary) \scriptsize{mon. = monitored, con. = controlled, der. = derived, decl. = macro rule declarations, rules = number of rules (inluding nested)}} 
	\label{tab:dimension} 
\end{minipage}\,\,\,\,%
\begin{minipage}[t]{0.44\textwidth}\vspace{0pt}%
\begin{lstlisting}[frame=lines, language=AsmetaL,
	breaklines=true,
	columns=fullflexible, label={code:MVM00}, caption={MVMController00 main rule}]
	main rule r_Main =
	par
	if state = STARTUP then	r_startup[]	endif
	if state = SELFTEST then r_selftest[] endif
	if state = VENTILATIONOFF then r_ventilationoff[] endif
	if state = PCV_STATE then r_runPCV[] endif
	if state = PSV_STATE then r_runPSV[] endif
	endpar
\end{lstlisting}
\end{minipage}
\end{figure*}

\subsection{First model: MVMController00}
The first model introduces the operation phases of the MVM controller. At the end of startup and self-test, the ventilator goes in the ventilation off state. Afterward,  on the basis of the user request, it can go to one of the two operation modes: PCV or PSV.
Code~\ref{code:MVM00} shows the main rule of the controller in the first model. It specifies the transitions among the MVM states by setting the value of the \code{state} variable (initialized at the \code{STARTUP} value). Depending on the \code{state} value, the corresponding rule is executed.

\begin{figure*}[tb]
	\centering
	\includegraphics[width=\linewidth]{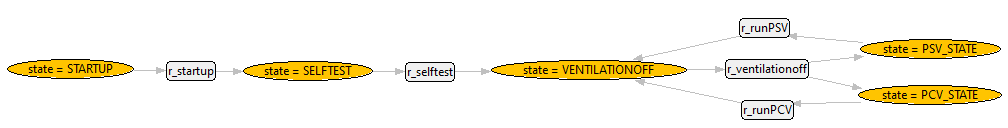}
	\caption{MVM state diagram}
	\label{fig:semVisMVM00}
\end{figure*}

The semantic visualization of the model, and in particular of the main rule, is shown in Fig.~\ref{fig:semVisMVM00}. It represents the MVM operation in terms of a control state machine: the value of the variable \code{state} is used as state mode to determine machine states.

\subsection{Second model: MVMController01}
The second model refines the inspiration and expiration phases in PCV and PSV mode.
Code~\ref{code:MVM01PCV} shows the refinement of the rule \code{r\_runPCV}, which here calls rules for the inspiration \code{r\_runPCVInsp} (line~\ref{line:MVM01PCVInsp}) and the expiration \code{r\_runPCVExp} (line~\ref{line:MVM01PCVExp}) rules.

In PCV mode, the transition between inspiration and expiration is determined by the duration of each phase decided by the physician (when timers \code{timerInspirationDurPCV}, in case of inspiration, and \code{timerExpirationDurPCV}, in case of expiration, expire).
When the inspiration time is passed (line~\ref{line:MVM01PCVinspirationExpired}), the controller goes to the PCV expiration phase (line~\ref{line:MVM01expPCV}). If the physician has required (by setting the value of the monitored function \code{respirationMode}) to move to PSV mode the machine changes the state from PCV to PSV and executes the rule \code{r\_PSVStartExp} (line~\ref{line:MVM01PCVTOPSV}). If a stop request (by the monitored function \code{stopRequested}) is received during the inspiration phase, it is stored (in \code{stopVentilation}) and will be executed in the expiration phase (line~\ref{line:MVM01PCVstopRequest}). When in expiration, if no stop request is received, the ventilator moves to PCV inspiration when expiration duration expires (line~\ref{line:MVM01PCVexpirationExpired}).

In PSV mode (see Code~\ref{code:MVM01PSV}), the transition from inspiration to expiration happens when the airflow drops a defined threshold (\code{flowDropPSV} holds, line~\ref{line:MVM01PSVinspirationExpired}) after a minimum inspiration time (\code{timerMin\-InspTimePSV} expires), or when the maximum  inspiration time set by the doctor is expired. The opposite transition occurs after a minimum expiration time (\code{timerMin\-ExpTimePSV} expires, line~\ref{line:MVM01expPSV}). The transition to ventilation off is allowed only from the expiration phase regardless of the stop command is received. Moreover, the physician can change from PSV to PCV and without interrupting the ventilation when in expiration phase (line~\ref{line:MVM01PSVTOPCV}).

Depending on the ventilator state, the input (\code{iValve}) and output  (\code{oValve}) valves are in the following position: input valve is closed and output valve is open when the ventilator is not running or in the expiration phase, input valve is open and output valve is closed when it is in inspiration phase.

\begin{figure*}[t]
\centering
	\begin{minipage}[t]{0.48\textwidth}
		\begin{lstlisting}[frame=lines, language=AsmetaL, numbers=left,
	stepnumber=1,
			columns=fullflexible, label={code:MVM01PCV}, caption={MVMController01 PCV}]
rule r_runPCV =
	par
		if phase = INSPIRATION then r_runPCVInsp[] endif
		if phase = EXPIRATION then r_runPCVExp[] endif
	endpar
rule r_runPCVInsp =	|\label{line:MVM01PCVInsp}|
	par
		if not stopVentilation then |\label{line:MVM01PCVstopRequest}|
			if stopRequested then stopVentilation := true endif
		endif
		if expired(timerInspirationDurPCV)	then |\label{line:MVM01PCVinspirationExpired}|
			par
				if respirationMode = PCV then
					r_PCVStartExp[] endif |\label{line:MVM01expPCV}|
				if respirationMode = PSV then |\label{line:MVM01PCVTOPSV}|
					par
						state := PSV_STATE
						r_PSVStartExp[]
					endpar
				endif
			endpar
		endif
	endpar
rule r_runPCVExp =		|\label{line:MVM01PCVExp}|
	if stopVentilation then r_stopVent[]
	else if stopRequested then r_stopVent[]
	else if expired(timerExpirationDurPCV) then  |\label{line:MVM01PCVexpirationExpired}|
		r_PCVStartInsp[]
	endif endif 	endif
rule r_PCVStartInsp =
	par
		phase := EXPIRATION
		iValve := CLOSED
		oValve := OPEN
		r_reset_timer[timerInspirationDurPCV]
	endpar
\end{lstlisting}	\end{minipage}\,\,\,\,%
\begin{minipage}[t]{0.48\textwidth}
		\begin{lstlisting}[frame=lines, language=AsmetaL,
		columns=fullflexible, label={code:MVM01PSV}, caption={MVMController01 PSV}]
rule r_runPSV =
	par
		if phase = INSPIRATION then r_runPSVInsp[] endif
		if phase = EXPIRATION then r_runPSVExp[] endif
	endpar
rule r_runPSVInsp =		|\label{line:MVM01PSVInsp}|
	par
		if not stopVentilation then |\label{line:MVM01PSVstopRequest}|
			if stopRequested then stopVentilation := true endif
		endif
		if (expired(timerMinInspTimePSV) and flowDropPSV)
			or expired(timerMaxInspTimePSV) then  |\label{line:MVM01PSVinspirationExpired}|
			r_PSVStartExp[]
		endif
	endpar
rule r_runPSVExp =		|\label{line:MVM01PSVExp}|
	if stopVentilation then r_stopVent[]
	else if stopRequested then r_stopVent[]
	else if expired(timerMinExpTimePSV) then |\label{line:MVM01expPSV}|
		par
			if respirationMode = PCV then |\label{line:MVM01PSVTOPCV}|
				par
					state := PCV_STATE
					r_PCVStartInsp[]
				endpar
			endif
			if respirationMode = PSV then	r_PSVStartInsp[]	endif
		endpar	endif	endif	endif
rule r_PSVStartInsp =
	par
		phase := EXPIRATION
		iValve := CLOSED
		oValve := OPEN
		r_reset_timer[timerMinExpTimePSV]
		r_reset_timer[timerMaxInspTimePSV]
	endpar
	\end{lstlisting}
	\end{minipage}
\end{figure*}

\begin{figure*}[!ht]
	\centering
	\includegraphics[width=0.7\linewidth]{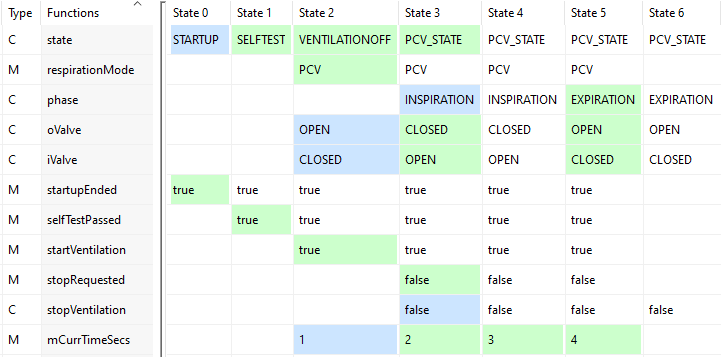}
	\caption{PCV animation example}
	\label{fig:animation}
\end{figure*}

\paragraph{Model validation.}
While modeling the ventilator, we have performed validation activities: animation   (simulation traces), model review, and validation.
An example of animation is reported in Fig.~\ref{fig:animation}, where the ventilator, after performing startup and self-test, is in the ventilation off state. As expected, the input valve is closed and the output valve is opened. When the start PCV command is sent to the ventilator, the PCV mode starts from the inspiration phase, and the valves are moved to the expected position: the input valve is opened and the output valve is closed. After the inspiration duration, the ventilator is in the expiration phase, the input valve is closed while the output valve is opened.

For the validation phase, we have written scenarios to check, whenever it is needed, that the desired behavior is captured by the model. Code~\ref{code:scenarioRef1} reports a scenario where at each \code{\textbf{step}} of the machine we check the ventilator state (\code{\textbf{check} state}) and the position of the input (\code{\textbf{check} iValve}) and output (\code{\textbf{check} oValve}) valves, given the inputs received (\code{\textbf{set}}).

\begin{figure*}[t]
	\centering
	\begin{minipage}[t]{0.31\textwidth}%
		\begin{lstlisting}[frame=lines, language=Avalla,
			breaklines=true,
			columns=fullflexible]
check state = STARTUP;
set startupEnded := true;
step
check state = SELFTEST;
set selfTestPassed := true;
step
check state = VENTILATIONOFF;
set startVentilation := true;
set respirationMode := PCV;
step
		\end{lstlisting}
	\end{minipage}\,\,\,\,%
	\begin{minipage}[t]{0.31\textwidth}%
		\begin{lstlisting}[frame=lines, language=Avalla,
			breaklines=true,
			columns=fullflexible]
check state = PCV_STATE;
check oValve = CLOSED;
check phase = INSPIRATION;
check iValve = OPEN;
step
check state = PCV_STATE;
check oValve = CLOSED;
check phase = INSPIRATION;
check iValve = OPEN;
step
		\end{lstlisting}
	\end{minipage}\,\,\,\,%
\begin{minipage}[t]{0.31\textwidth}%
	\begin{lstlisting}[frame=lines, language=Avalla,
		breaklines=true,
		columns=fullflexible]
check state = PCV_STATE;
check oValve = OPEN;
check phase = EXPIRATION;
check iValve = CLOSED;
step
check state = PCV_STATE;
check oValve = OPEN;
check phase = EXPIRATION;
check iValve = CLOSED;
check stopVentilation = false;
	\end{lstlisting}
\end{minipage}
	\vspace{-10pt}
	\begin{lstlisting}[caption={PCV scenario example},
		label={code:scenarioRef1},
		captionpos=t]
	\end{lstlisting}
	\vspace{-0.5cm}
\end{figure*}

\paragraph{Model verification.}
Using the model checker, we have verified the following safety properties:
\begin{compactitem}
	\item Valves are never both open or closed at the same time: \begin{lstlisting}[language=AsmetaL,float=false,basicstyle=\small]
	LTLSPEC not f(iValve=oValve)
	\end{lstlisting}
	\item When ventilation is off, the output valve is open and the input valve is closed \begin{lstlisting}[language=AsmetaL,float=false,basicstyle=\small]
	LTLSPEC g(state=VENTILATIONOFF implies (iValve=CLOSED and oValve=OPEN))
	\end{lstlisting}
\end{compactitem}

These two properties are crucial for the safety of the ventilator. The former guarantees that the patient can not choke, while the latter assures that the system is fail-safe since, when the ventilator is off, the relief valve allows the patient to breathe.
 
\subsection{Third model: MVMController02}
At the end of the inspiration phase, the physician can perform an inspiratory pause or recruitment maneuver, and the expiratory pause is allowed after the expiration phase.
This has been modeled as shown in Code~\ref{code:MVM02PCV} for PCV mode and Code~\ref{code:MVM02PSV} for PSV mode, which respectively extend the behavior of the MVMController01 as shown in  Code~\ref{code:MVM01PCV} and Code~\ref{code:MVM01PSV}.

After inspiration, if an inspiratory pause is required (monitored function \code{cmdInPause} holds, Code~\ref{code:MVM02PCV} line~\ref{line:MVM02PCVInPause} and Code~\ref{code:MVM02PSV} line~\ref{line:MVM02PSVInPause}) the valves are both closed for the entire pause duration.
If inspiratory pause is not required, the doctor can select the recruitment maneuver (by setting \code{cmdRm}, Code~\ref{code:MVM02PCV} line~\ref{line:MVM02PCVRm} and Code~\ref{code:MVM02PSV} line~\ref{line:MVM02PSVRm}) and the lungs are filled with oxygen and medical air.

An expiratory pause can be performed upon the doctor's request after the expiration phase (monitored function \code{cmdExPause} holds, Code~\ref{code:MVM02PCV} line~\ref{line:MVM02PCVExPause}, and Code~\ref{code:MVM02PSV} line~\ref{line:MVM02PSVExPause}).

In the expiratory and inspiratory pause states, 
the input and output valves are both closed, while in the recruitment maneuver the output valve is closed and the input valve is opened to allow the air flowing in the alveoli.
Moreover, as shown in Code~\ref{code:MVM02PSV}, when PSV is running and the ventilator does not detect a new breath within apnea lag (timer \code{timerApneaLag} expires, line~\ref{line:MVM02PSVApnea}), the ventilator automatically changes to PCV mode starting from the inspiration phase (line~\ref{line:MVM02Apnea}).

\begin{figure*}[t]
	\centering
	\begin{minipage}[t]{0.48\textwidth}
		\begin{lstlisting}[frame=lines, language=AsmetaL, numbers=left,
			stepnumber=1,
			columns=fullflexible, label={code:MVM02PCV}, caption={MVMController02 PCV}]
rule r_runPCV =
	par		...
		if phase = INPAUSE then	r_runInPause[] endif
		if phase = RM then	r_runRm[]	endif
		if phase = EXPAUSE then	r_runExPause[]	endif
	endpar
rule r_runPCVInsp =	|\label{line:MVM02PCVInsp}|...
	if expired(timerInspirationDurPCV)	then
	par
			if respirationMode = PCV then
				if cmdInPause then	 |\label{line:MVM02PCVInPause}|
					r_InPause[]
				else
					if cmdRm then		r_rm[] |\label{line:MVM02PCVRm}|
					else	r_PCVStartExp[]
				endif
			endif endif
			if respirationMode = PSV then
				par
					state := PSV_STATE
					r_PSVStartExp[]
					r_resetApneaBackup[]
				endpar
			endif
		endpar
	endif ...
rule r_runPCVExp =		|\label{line:MVM02PCVExp}|...
	if expired(timerExpirationDurPCV) then
		if cmdExPause then |\label{line:MVM02PCVExPause}|
			r_exPause[]
		else
			r_PCVStartInsp[]
		endif
	endif 	 ...

	\end{lstlisting}	\end{minipage}\,\,\,\,%
	\begin{minipage}[t]{0.48\textwidth}
		\begin{lstlisting}[frame=lines, language=AsmetaL,
			columns=fullflexible, label={code:MVM02PSV}, caption={MVMController02 PSV}]
rule r_runPSV =
	par		...
		if phase = INPAUSE then	r_runInPause[] endif
		if phase = RM then	r_runRm[]	endif
		if phase = EXPAUSE then	r_runExPause[]	endif
	endpar
rule r_runPSVInsp =		|\label{line:MVM02PSVInsp}|...
	if (expired(timerMinInspTimePSV) and flowDropPSV)
		or expired(timerMaxInspTimePSV) then
	if cmdInPause then	r_InPause[] |\label{line:MVM02PSVInPause}|
	else if cmdRm then	r_rm[] |\label{line:MVM02PSVRm}|
	else	r_PSVStartExp[]	endif endif
	endif ...
rule r_runPSVExp =...
	if expired(timerApneaLag) then	r_runApnea[] |\label{line:MVM02PSVApnea}|
	else 	if expired(timerMinExpTimePSV) then
		par
			if respirationMode = PCV then
				par
					state := PCV_STATE
					r_PCVStartInsp[]
				endpar
			endif
			if respirationMode = PSV then
				if cmdExPause then	r_ExPause[]		endif  |\label{line:MVM02PSVExPause}|
			endif
		endpar
	endif	endif ...
rule r_runApnea =
	par
		state := PCV_STATE
		r_PCVStartInsp[] |\label{line:MVM02Apnea}|
		apneaBackupMode := true
	endpar
		\end{lstlisting}
	\end{minipage}
\end{figure*}

\paragraph{Model validation and verification.}
Model validation activities have been performed also at this level.
Considering the properties verified in the previous refinement step, the property that states ``valves are never closed at the same time" does not hold anymore. Indeed, when the ventilator is in pause the valves are both closed as guaranteed by this property:
\begin{lstlisting}[language=AsmetaL,float=false,basicstyle=\small]
LTLSPEC g(((phase=INPAUSE or phase=EXPAUSE) and (state = PCV_STATE or state = PSV_STATE)) implies
 (iValve=CLOSED and oValve=CLOSED))
\end{lstlisting}
To ensure that the valves are never both closed outside inspiratory and expiratory pause, we have verified the following property, in which valves are both closed if the ventilator is not in inspiration, expiration, recruitment maneuver, ventilation off, startup, or selftest.
\begin{lstlisting}[language=AsmetaL,float=false,basicstyle=\footnotesize]
LTLSPEC g((iValve=CLOSED and oValve=CLOSED) implies ((not ((phase=INSPIRATION or 
phase=EXPIRATION or phase=RM) and (state = PCV_STATE or state = PSV_STATE)))) or 
(not (state = VENTILATIONOFF or state = STARTUP or state = SELFTEST)))
\end{lstlisting}

\subsection{Fourth model: MVMController03}
In the last model, we have introduced the transition from inspiration to expiration and vice versa depending on the pressure changes due to spontaneous breathing. The new behavior has been modeled by extending the rules \code{r\_runPCVInsp} and \code{r\_runPCVExp} as shown in Code~\ref{code:MVM03PCV}, and rules \code{r\_runPSVInsp} and \code{r\_runPSVExp} as shown in Code~\ref{code:MVM03PSV}.

When the ventilator is in expiration (Code~\ref{code:MVM03PCV} line~\ref{line:MVM03PCVExp} and Code~\ref{code:MVM03PSV} line~\ref{line:MVM03PSVExp}) and it detects after an instant of time (a \emph{trigger window} here modeled as the expiration of the timer \code{timerTriggerWindowDelay}) a sudden drop in pressure below the inhale trigger sensitivity threshold  (monitored function \code{dropPAW\_ITS} holds, Code~\ref{code:MVM03PCV} line~\ref{line:MVM03PCVDropPAW} and Code~\ref{code:MVM03PSV} line~\ref{line:MVM03PSVDropPAW}), the ventilator directly moves to the inspiration phase.

The transition from inspiration to expiration is automatically performed when the pressure goes beyond the maximum threshold set by the doctor (monitored function \code{pawGTMaxPinsp} holds, Code~\ref{code:MVM03PCV} line~\ref{line:MVM03PCVMaxPinsp} and Code~\ref{code:MVM03PSV} line~\ref{line:MVM03PSVMaxPinsp}).
At this third refinement level, we have performed the validation and verification activities as done for the previous levels.

\begin{figure*}[t]
	\centering
	\begin{minipage}[t]{0.48\textwidth}
		\begin{lstlisting}[frame=lines, language=AsmetaL, numbers=left,
			stepnumber=1,
			columns=fullflexible, label={code:MVM03PCV}, caption={MVMController03 PCV}]
rule r_runPCVInsp =	|\label{line:MVM03PCVInsp}|
...
	if expired(timerInspirationDurPCV)	then
		...
	else if pawGTMaxPinsp then |\label{line:MVM03PCVMaxPinsp}|
		r_PCVStartExp[]
	endif endif
rule r_runPCVExp =		|\label{line:MVM03PCVExp}|
...
	if expired(timerExpirationDurPCV) then
		...
	else if expired(timerTriggerWindowDelay) and dropPAW_ITS then
		r_PCVStartInsp[] |\label{line:MVM03PCVDropPAW}|
	endif endif ...
	\end{lstlisting}	\end{minipage}\,\,\,\,%
	\begin{minipage}[t]{0.48\textwidth}
		\begin{lstlisting}[frame=lines, language=AsmetaL,
			columns=fullflexible, label={code:MVM03PSV}, caption={MVMController03 PSV}]
rule r_runPSVInsp =		|\label{line:MVM03PSVInsp}|
...
	if (expired(timerMinInspTimePSV) and flowDropPSV) or
	 expired(timerMaxInspTimePSV) then
		...
	else if pawGTMaxPinsp then |\label{line:MVM03PSVMaxPinsp}|
		r_PCVStartExp[]
	endif endif
rule r_runPSVExp =		|\label{line:MVM03PSVExp}|
...
	if expired(timerTriggerWindowDelay) and dropPAW_ITS then  |\label{line:MVM03PSVDropPAW}|
		r_PSVStartInsp[]
	else if expired(timerApneaLag) then
		...
		\end{lstlisting}
	\end{minipage}
\end{figure*}

\section{C++ automatic code generation and unit testing}\label{sec:codegen} 

After having modeled the mechanical ventilator controller with \asmetaL and verified the specification, we have automatically generated the \cpp code using the \asmtocode tool starting from the last model, \code{MVMController03.asm}. The tool generates two different files, \code{MVMController03.h} and \code{MVMController03.cpp}, which contain the translation of the ASM model as a \cpp class.
During the \cpp code generation, each ASM rule is translated into a class method. An example of the content of the \code{MVMController03.cpp} file is reported in Code~\ref{code:cppFile}, which contains the \cpp translation of the two rules previously shown in Code~\ref{code:MVM03PSV}.
\begin{figure*}[t]
	\centering
	\begin{minipage}[t]{0.48\textwidth}
		\begin{lstlisting}[frame=lines, language=C++, numbers=left,
			stepnumber=1,
			breaklines=true,
			columns=fullflexible, tabsize=2, columns=fullflexible, basicstyle=\scriptsize, captionpos=b, morekeywords={REQUIRE, TEST_CASE}]
[...]
void MVMController03::r_runPCVInsp(){
	if (!stopVentilation[0]){ ... }
	if (expired(timerInspirationDurPCV)){
		if ((respirationMode == PCV)){
			if (cmdInPause){
				r_InPause();
			} else if (cmdRm){
				r_rm();
			} else {
				r_PCVStartExp();
			}
		}
	} else if (pawGTMaxPinsp)
		r_PCVStartExp();
}
		\end{lstlisting}	\end{minipage}\,\,\,\,%
		\begin{minipage}[t]{0.48\textwidth}
			\begin{lstlisting}[frame=lines, language=C++,
				breaklines=true,
				columns=fullflexible, tabsize=2, columns=fullflexible, basicstyle=\scriptsize, captionpos=b, morekeywords={REQUIRE, TEST_CASE}]
void MVMController03::r_runPCVExp(){
	if (stopVentilation[0]){
		r_stopVent();
	} else if (stopRequested){
		r_stopVent();
	} else if (expired(timerExpirationDurPCV)){
		if (cmdExPause){
			r_exPause();
		}else{
			r_PCVStartInsp();
		}
	} else if (expired(timerTriggerWindowDelay) & dropPAW_ITS){
		r_PCVStartInsp();
	}
}
[...]
		\end{lstlisting}
	\end{minipage}
	\begin{lstlisting}[captionpos=t, label={code:cppFile},caption={Example of the \code{MVMController03.cpp} file}]
	\end{lstlisting}
\end{figure*}

Besides the source code implementing the ASM in \cpp, we have automatically generated the unit tests in \cpp for that code.
The automatic test generation is performed by the {\tt ATGT} tool, which can exploit both the counterexamples given by the model checker and random traces generated by the \asmeta simulator.
{\tt ATGT} produces abstract tests as sequences of ASM states. These tests are then concretized using the Catch2 framework\footnote{\url{https://github.com/catchorg/Catch2}} and used for performing unit testing on the \cpp code (see Code~\ref{code:catchTest}).
Each test contains the declaration of the object under test (the \code{mvmcontroller03} in our case). Then, the following steps are repeated:
\begin{inparaenum}[(I)]
	\item the monitored functions are set (such as \code{startupEnded} in Code~\ref{code:catchTest}),
	\item the possible controlled functions, which need to be initialized to the same value of the monitored ones, are set using the \code{initControlledWithMonitored()} method,
	\item the main rule \code{r\_Main()} is executed,
	\item the update set is fired using the \code{fireUpdateSet()} method,
	\item the values of controlled functions are checked against the expected ones (for instance, \code{state} in Code~\ref{code:catchTest}).
\end{inparaenum}
\begin{figure*}[t]
	\centering
	\begin{minipage}[t]{0.48\textwidth}
		\begin{lstlisting}[frame=lines, language=C++,
		breaklines=true,
		columns=fullflexible, tabsize=2, columns=fullflexible, basicstyle=\scriptsize, captionpos=b, morekeywords={REQUIRE, TEST_CASE}]
[...]
TEST_CASE( "my_test_0", "my_test_0"){
	// instance of the SUT
		MVMController03 mvmcontroller03;
	// init controlled with monitored term
		mvmcontroller03.initControlledWithMonitored();
	// check controlled variables
		REQUIRE(mvmcontroller03.state[0] == STARTUP);
	// set monitored variables
		mvmcontroller03.startupEnded = true;
		mvmcontroller03.mCurrTimeSecs=1;
	\end{lstlisting}	\end{minipage}\,\,\,\,%
	\begin{minipage}[t]{0.48\textwidth}
		\begin{lstlisting}[frame=lines, language=C++,
				breaklines=true,
				columns=fullflexible, tabsize=2, columns=fullflexible, basicstyle=\scriptsize, captionpos=b, morekeywords={REQUIRE, TEST_CASE}]
// call main rule
	mvmcontroller03.r_Main();
	mvmcontroller03.fireUpdateSet();
// check controlled variables
	REQUIRE(mvmcontroller03.state[0] == SELFTEST);
// set monitored variables
	mvmcontroller03.mCurrTimeSecs=2;
// call main rule
	mvmcontroller03.r_Main();
	mvmcontroller03.fireUpdateSet();
[...] }
		\end{lstlisting}
	\end{minipage}
	\begin{lstlisting}[captionpos=t, label={code:catchTest},caption={Example of a Catch2 test case}]
	\end{lstlisting}
\end{figure*}
For the MVM controller, the testing process requires the simulation of the time as well: the time is represented as a monitored function whose value is incremented by one second at each step. As shown in Code \ref{code:catchTest}, \emph{mCurrTimeSecs} takes value initially 1 and is then incremented by 1 at each step.

In this case study, we have used  the random test generation based on the use of the simulator. The user can configure the test generation process by choosing the number of tests to be generated and the number of steps each.
We have generated 50 tests of 50 steps each and we have verified the compliance of the \cpp code with the \code{MVMController03.asm} model. We have measured the code coverage and found that with the automatically generated tests we have covered $100\%$ of code  in terms of instructions of the \code{MVMController03.cpp} and \code{MVMController03.h} classes generated from the model. 
We have investigated why we have been able to completely cover the generated code and found that the tests, although they are randomly generated, cover every possible ASM transition, and the tool \asmtocode does not generate dead code.

\section{Code deployment on Arduino}\label{sec:arduino}

\begin{figure}
	\centering
	\includegraphics[width=0.7\linewidth]{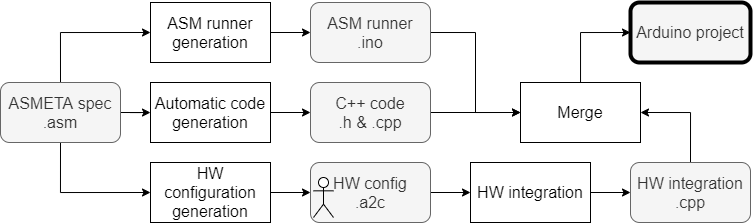}
	\caption{Arduino code generation process}
	\label{fig:cppgeneration}
\end{figure}

Exploiting the \asmtocode tool, \asmeta allows the deployment of the \cpp code as an Arduino sketch too~\cite{Bonfanti2017}. The entire process is depicted in Fig.~\ref{fig:cppgeneration}.
After the generation of the \texttt{.h} and \texttt{.cpp} files as explained in Sect.~\ref{sec:codegen}, \asmtocode automatically generates an \texttt{.a2c} file.
This configuration file is used for binding each ASM function to
an Arduino physical pin.
It must be completed by the user who has to insert the correspondence between Arduino physical pins and functions defined in the ASM model. For example, in Code~\ref{code:a2c}, input and output valves (\code{iValve} and \code{oValve}) are mapped on digital output pins, while the monitored functions are used to set if the current phase is finished or not (e.g. \code{startupEnded} and \code{selfTestPassed}) are read using digital input pins.

\begin{figure*}[t]
	\centering
	\begin{minipage}[t]{0.31\textwidth}%
		\begin{lstlisting}[frame=lines, language=JSON,
			breaklines=true,
			columns=fullflexible, tabsize=2, basicstyle=\scriptsize]
{
"arduinoVersion": "UNO",
"stepTime": 0,
"bindings": [
	{
		"mode": "DIGITALOUT",
		"function": "iValve",
		"pin": "D8"
	},
		\end{lstlisting}
	\end{minipage}\,\,\,\,%
	\begin{minipage}[t]{0.31\textwidth}%
		\begin{lstlisting}[frame=lines, language=JSON,
			breaklines=true,
			columns=fullflexible, tabsize=2, basicstyle=\scriptsize]
	{
		"mode": "DIGITALOUT",
		"function": "oValve",
		"pin": "D7"
	}, {
		"mode": "DIGITALIN",
		"function": "startupEnded",
		"pin": "A5"
	},
		\end{lstlisting}
	\end{minipage}\,\,\,\,%
\begin{minipage}[t]{0.31\textwidth}%
\begin{lstlisting}[frame=lines, language=JSON,
	breaklines=true,
	columns=fullflexible, tabsize=2, basicstyle=\scriptsize]
	{
		"mode": "DIGITALIN",
		"function": "selfTestPassed",
		"pin": "A4"
	},
	
	[...]
	
}
\end{lstlisting}
\end{minipage}
	\vspace{-10pt}
	\begin{lstlisting}[caption={Example of the a2c configuration file},
		label={code:a2c},
		captionpos=t]
	\end{lstlisting}
	\vspace{-0.5cm}
\end{figure*}

Having completed the \texttt{.a2c} file with the correct mappings, \asmtocode generates two additional files: the \texttt{hw.cpp} and the \texttt{.ino}. The former (see Code~\ref{code:hw}) implements the hardware-specific methods, i.e. those related to the reading of inputs (\code{getInputs()} method in Code~\ref{code:hw}) and writing of outputs (\code{setOutputs()} method in Code~\ref{code:hw}) through physical pins. The latter (see Code~\ref{code:ino}), contains the execution policy to run the ASM on Arduino and performs cyclically the following operations:
\begin{inparaenum}[(i)]
	\item \code{getInputs()} reads the inputs through digital and analog pins,
	\item \code{r\_Main()}, represents the main rule of the ASM and executes all the rules,
	\item \code{setOutputs()}, sends the output values through the physical Arduino pins to the output components,
	\item \code{fireUpdateSet()}, updates the values of controlled functions to be used in the next state.
\end{inparaenum}

\begin{figure*}[!ht]
\centering
\begin{minipage}{.49\textwidth}
	\centering
	\includegraphics[width=0.98\linewidth]{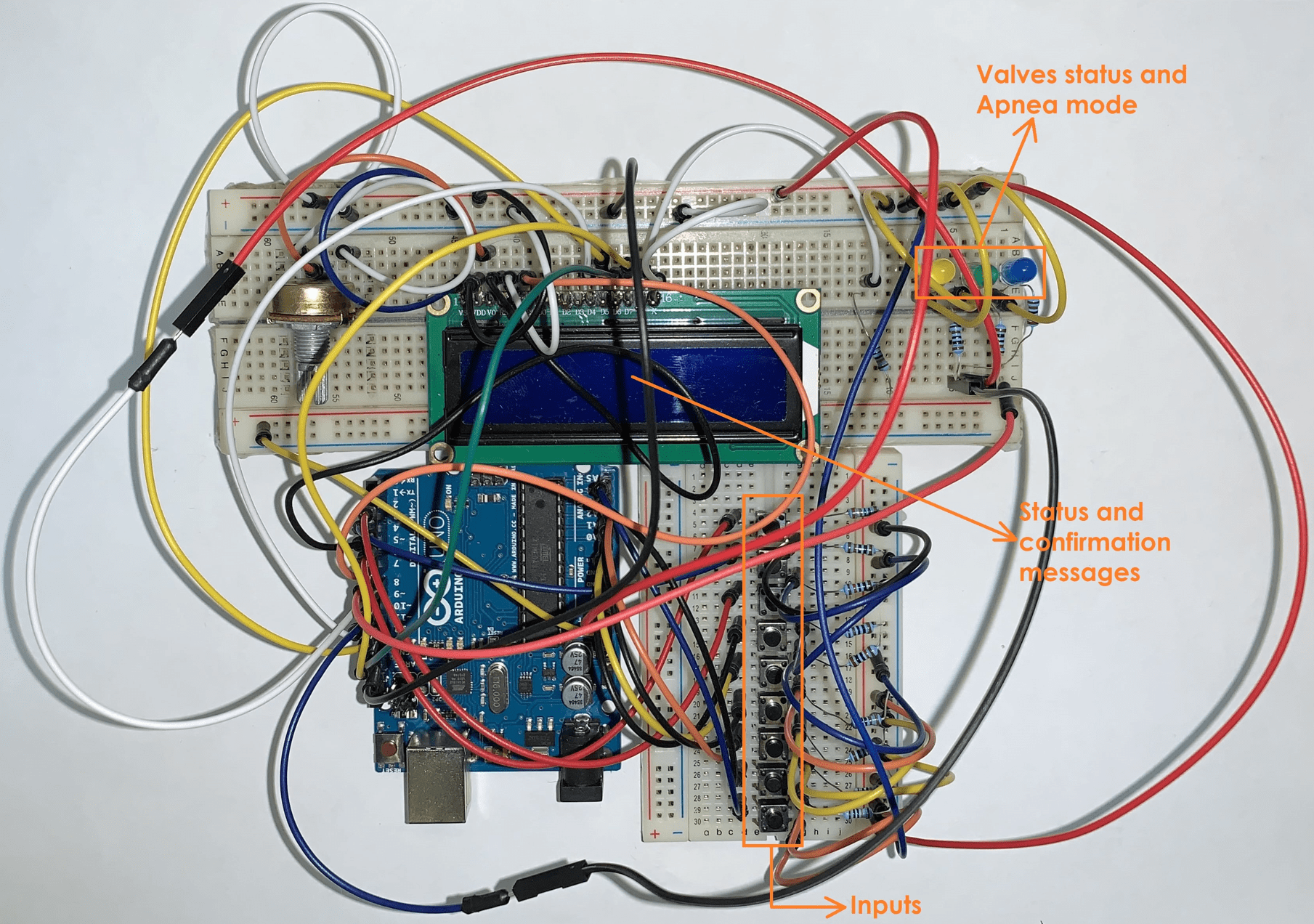}
	\caption{The Arduino version of the MVM}
	\label{fig:Arduino}
	
\end{minipage}
\begin{minipage}{.5\textwidth}
\centering
\includegraphics[width=.95\linewidth]{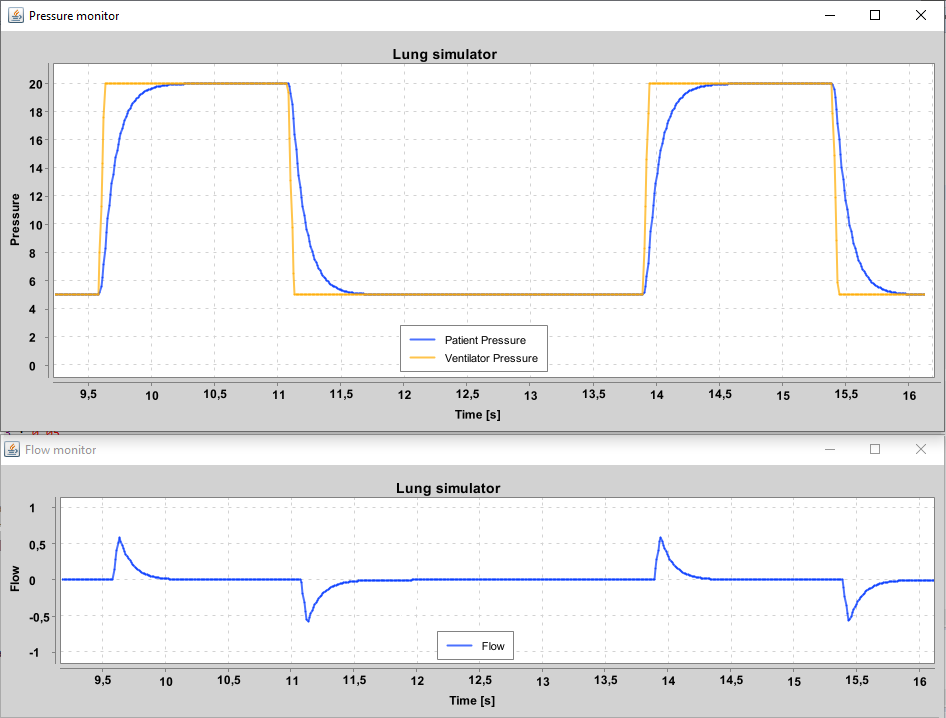}
\caption{Example of the breathing simulation}
\label{fig:breathingsimulator}
\end{minipage}
\end{figure*}

When modeling time we have exploited the \code{TimeLibrary} for time management. However, Arduino has very limited support for time and timers. We decided to use the \texttt{millis} instruction to implement temporized operations. \texttt{millis} returns the number of milliseconds passed since the Arduino board began running the current program and we have translated all the operations over the timers in terms of this function.
\begin{figure*}[t]
	\centering
	\begin{minipage}[t]{0.48\textwidth}%
		\begin{lstlisting}[frame=lines, language=C++,
			breaklines=true,
			columns=fullflexible, tabsize=2, basicstyle=\scriptsize]
#include "MVMController03.h"

void MVMController03::getInputs(){
	startupEnded = (digitalRead(A5) == HIGH);
	selfTestPassed = (digitalRead(A4) == HIGH);
	[...]
}
void MVMController03::setOutputs(){
	if (iValve[0] != iValve[1]){
		if(iValve == OPEN)
			digitalWrite(8, LOW);
			\end{lstlisting}
		\end{minipage}\,\,\,\,%
		\begin{minipage}[t]{0.48\textwidth}%
			\begin{lstlisting}[frame=lines, language=C++,
				breaklines=true,
				columns=fullflexible, tabsize=2, basicstyle=\scriptsize]
	else
		digitalWrite(8, HIGH);
	}
	if (oValve[0] != oValve[1]){
		if(oValve == OPEN)
			digitalWrite(7, LOW);
		else
			digitalWrite(7, HIGH);
	}
	[...]
}
		\end{lstlisting}
	\end{minipage}
	\vspace{-10pt}
	\begin{lstlisting}[caption={Extract of the \texttt{hw.cpp} file containing hardware-specific functions},
		label={code:hw},
		captionpos=t]
	\end{lstlisting}
	\vspace{-0.5cm}
\end{figure*}
\begin{figure*}[t]
	\centering
	\begin{minipage}[t]{0.48\textwidth}%
		\begin{lstlisting}[frame=lines, language=C++,
			breaklines=true,
			columns=fullflexible, tabsize=2, basicstyle=\scriptsize]
#include"MVMController03.h"

void setup(){
	pinMode(8, OUTPUT);
	// ... set all the other outputs
	pinMode(A2, INPUT);
	// ... set all the other inputs
}
\end{lstlisting}
\end{minipage}\,\,\,\,%
\begin{minipage}[t]{0.48\textwidth}%
\begin{lstlisting}[frame=lines, language=C++,
					breaklines=true,
					columns=fullflexible, tabsize=2, basicstyle=\scriptsize]
MVMController03 mvmcontroller03;

void loop(){
	mvmcontroller03.getInputs();
	mvmcontroller03.r_Main();
	mvmcontroller03.setOutputs();
	mvmcontroller03.fireUpdateSet();
}
		\end{lstlisting}
	\end{minipage}
	\vspace{-10pt}
	\begin{lstlisting}[caption={Example of the \texttt{.ino} file containing the implementation of the ASM execution},
		label={code:ino},
		captionpos=t]
	\end{lstlisting}
	\vspace{-0.5cm}
\end{figure*}
We have chosen the hardware to be used in our Arduino-based implementation of the simplified MVM as follows (see Fig.~\ref{fig:Arduino}):
\begin{compactitem}
	\item Arduino Uno, that executes the state machine;
	\item 3 LEDs used to communicate the status of input and output valves and the apnea alarm; one 1602 LCD display, which shows the current state;
	\item 9 buttons, which simulates all the monitored functions contained in the ASM, namely \code{dropPAW\_ITS}, \code{pawGTMaxPinsp}, \code{cmdRm}, \code{cmdInPause}, \code{cmdExPause}, \code{flowDropPSV}, \code{respirationMode}, \code{stopRequested}, \code{startupEnded}, \code{selfTestPassed}, and \code{startVentilation}. %\red{S:quali?}
	They represent both user inputs and external breathing events.
 \end{compactitem}
 
\begin{figure*}[!ht]
\centering
\includegraphics[width=0.4\linewidth]{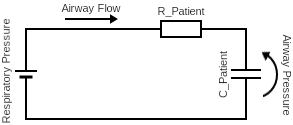}
\caption{RC equivalent circuit of a human lung}
\label{fig:RC}
\end{figure*}

One can play with the MVM on Arduino by pressing the buttons and observing the status of the MVM. However, as it is, it provides a very limited user interface. To enable a more meaningful user experience, we have developed a custom-made Java \emph{breathing simulator}. It is based on a simple lung model~\cite{Campbell1963THEEA}
whose electrical equivalent circuit is shown in Fig. \ref{fig:RC}. The breathing simulator allows the simulation of different patients, by setting the right values for resistance and compliance of their lungs, it can provide the measure of the pressure and flow of air in the patient lungs, and moreover, it can simulate events related to the patient's breathing like a drop in the pressure when the patient starts.
We connected the software simulator to the MVM on the Arduino using a serial port - so the MVM can communicate the status of the valves and therefore set the pressure of the ventilator and can read breathing events. An example of the breathing simulation is shown in Fig. \ref{fig:breathingsimulator}.


We have tested the Arduino version of the MVM and a video, showing the functioning in all the ventilation modes, can be found at the following link \url{https://youtu.be/a3fhqLpYVMI}.

\section{Discussion and Conclusions}\label{sec:discussion}

In this section, we try to address the concerns and answer the questions proposed by the organizers. We believe that applying \asmeta to a real case study, that has been successfully completed with our direct involvement but without the use of formal methods, can put us in the privileged position to better understand the strength and weaknesses points of \asmeta and what is still needed to make it applicable in the future. 
Regarding \emph{applicability}, \asmeta has never been applied 
in industrial projects, but we have used it in  industrial 
case studies provided in several contexts like in the medical sector~\cite{Bombarda2019,abz2016siSCICO2017}, automotive~\cite{ArcainiABZ2020}, and avionics~\cite{LGSjournalSTTT2017} to name a few.
Barriers to the wide adoption of a formal method do exist \cite{Garavel2020,Gleirscher2020} and thanks to this work we have a better vision of what we need to do in order to favor the use of \asmeta in an industrial setting.

We have always given great importance to \emph{automation} and our method is well supported by a set of tools (as explained in Sect. \ref{sec:asmeta}). In this case study, we did not suffer from the lack of a specific tool supporting a task. However, as explained below, our tools have different levels of maturity and sometimes the tool support has not been exhaustive.

For this work, \emph{integration} among tools and techniques has played an important role. We think that having a model on which several techniques can be applied, including V\&V, testing, and code generation is crucial for the use in the  context of critical systems. 
Some integration among \asmeta tools is still work-in-progress. For instance, the TimeLibrary is not directly supported by the test generator tool and the support for the model checker NuXmv is rather new and it has some problems. 
 
Another aspect to be considered is the integration of a formal approach into the engineering process. 
For MVM, as for some critical software lately \cite{Islam2020}, the software development process has been agile, with frequent changes in requirements that are then implemented in the code in a continuous integration way. \asmeta, as any Model-Driven Engineering approach, struggles in the integration with agile processes. For instance, direct modifications of the generated source code risk being lost if a change in the specification is performed and a new version of the code is generated.

For complex systems, modeling their entire behavior in one shot may be difficult. Thus, with the \asmeta \emph{methodology}, users can incrementally add details in models using the refinement technique (see Sect.~\ref{sec:vev}) or including already existing modules. Refinement and modularity are exploitable also when defining scenarios for model validation.
We envision the application of a formal process only to core components of the entire system. In the case of the MVM ventilator, for example, only the controller has been modeled by using \asmeta and the code obtained automatically from the specifications.  Although for other MVM components, like the alarm system, a formal methodology could be a good fit,  for other parts, such as graphical widgets or sensor drivers, 
formal specifications are possible but sometimes useless or even not suitable,
especially if the source code is already provided or it can be reused (for instance for sensor drivers). In any case, the methodology must foresee a phase in which the whole code 
(the one obtained through formal specification and the one handwritten)
is integrated (and tested) for building the complete system.  
In terms of the hardware platform, Arduino is a good choice when it comes to developing prototypes. However, for the real medical device, it should be replaced with a more robust one (MVM actually embeds a microcontroller that is Arduino code compatible).

One undoubted advantage regarding the \emph{usefulness} of using a formal approach is that the certification process, if required, is easier to complete. 
The main standard for medical software certification to be considered is the IEC 62304~\cite{IEC62304}. It does not prescribe a specific life cycle model but defines the process, activities, and tasks that the life cycle model has to follow.

In \cite{abz2016siSCICO2017}, we have identified how ASMs can be used to satisfy the process and how the activities prescribed by the standard can be mapped to activities performed by using our formal approach. After requirements are captured by models, the verification and validation activities are straightforward, and the desired system behavior is automatically transferred to the generated code. Moreover, the formal process allows for requirements traceability which is required by the certification. Wrt the MVM agile process used for deploying the device and that required a huge effort to provide and guarantee requirements traceability, here, the use of \asmeta formal approach makes traceability easier to demonstrate (e.g. by mapping requirements to rules).
When it comes to the usefulness in terms of the rapidness of code development, with the \asmeta framework and following the process described in this paper, we have been able to obtain the final code of the MVM controller in only a couple of days.
However, one must consider that when modeling with \asmeta we already knew requirements, having contributed to the device development.

One activity whose  \emph{usefulness}  is sometimes questioned is the unit test generation. After all, if the code is automatically generated from models, is there any need to validate it with tests derived from the same specification? We believe that there are at least three motivations:
\begin{inparaenum}[(i)]
\item the source code could be modified and the tests can check that the behavior as specified by the ASM is preserved;
\item the translation from ASM to \cpp itself must be validated, so unit tests check the conformance between the ASM and the \cpp code;
\item for certification purposes, source code must come with a test suite that provides confidence in its correctness.
	\end{inparaenum} 

Over the years, we have strived to improve \asmeta \emph{usability} \cite{afford19} by developing it as a set of  Eclipse plugins, and feedbacks from our students, as well as the application of the method to competitive case studies,  have been the basics for further improvements and extensions. However, it remains an academic tool whose stability and maintenance cannot reach the level often required in an industrial setting (and this was one of the reasons why we decided not to use it during the real MVM development).
Moreover, the improvements introduced in the years have made \asmeta similar to higher-level programming languages, as for other formal methods. Nevertheless, 
the pseudo-code style and freedom of abstraction in ASMs allow for capturing of requirements at a very high-level of abstraction in a form that is understandable by the stakeholders.

In conclusion, we have applied \asmeta to assemble a prototype of a system which we have contributed to develop and deploy in the recent past and we have been able to evaluate the feasibility of our formal process. Although we believe that there is still some work to be done in order to provide the necessary stability and maturity of the tools and of the process, we were able to carry on the development of the MVM controller from the requirement specification to the code.

\bibliographystyle{eptcs}
\bibliography{bibliography}

\end{document}